Polarimetry and the Long Awaited Superoutburst of BZ UMa

A. Price[1], J. Masiero[2], A. A. Henden[1], T. Vanmunster[3,1], D. Boyd[4,1], S. Dvorak[1], A. Oksanen[1], J. Shears[1], D. Starkey[1], A. Arminski[1], D. Collins[5], D. Durig[6], B. Harris[1], R. Koff[1], M. Koppelman[1], T. Krajci[1], M. Reszelski[1], M. Simonsen[1], T. Arranz[7], R. Tomlin[1], , D. Wells[1]

[1] American Association of Variable Star Observers

Clinton B. Ford Astronomical Data and Research Center,

49 Bay State Road,

Cambridge, MA 02138 USA

[2] Institute for Astronomy,

University of Hawaii,

2680 Woodlawn Dr.,

Honolulu, HI 96822 USA

[3] CBA Belgium Observatory

Walhostraat 1A

B-3401 Landen, Belgium

[4] British Astronomical Association,

Burlington House,

Piccadilly, London W1V 0DU UK

[5]Warren Wilson College,

PO BOX 9000,

Asheville, NC 28815

[6]The Cordell-Lorenz Observatory

The University of the South

Sewanee, TN 37383

[7]El Observatorio Astronómico

Las Pegueras

40470 Navas De Oro (Segovia), España.

Abstract

BZ UMa is a cataclysmic variable star whose specific classification has eluded researchers since its discovery in 1968. It has outburst and spectral properties consistent with both U Gem class dwarf novae and intermediate polars. We present new photometric and polarimetric measurements of recent outbursts, including the first detected superoutburst of the system. Statistical analysis of these and archival data from outbursts over the past 40 years present a case for BZ UMa as a non-magnetic, U Gem class, SU-UMa subclass dwarf novae.

Keyword(s): Stars

*Introduction.*

Cataclysmic variable (CV) stars are multiple star systems that suddenly, and often unpredictably, increase in brightness for a short time period. There are many sub classifications of CV systems, usually named after their prototypical stars (see Warner 2003 for a review). One of the most numerous subclasses are the U Gem type, SU-UMa subtype (UGSU) dwarf novae (Vogt 1980). In these systems a white dwarf accretes material from a less massive, orbiting secondary. This material collects into an accretion disc around the white dwarf and eventually releases energy, increasing the optical brightness of the system by many magnitudes. Eventually, the outburst ends in a few days to weeks and the system returns to its normal brightness. This cycle will repeat itself on the order of weeks to years. Superoutbursts of short period CVs exhibit superhumps, modulations of a few tenths of a magnitude with a periodicity usually a few percent longer than the orbital period of the star (Patterson 1998 Superhumps are likely caused by the precession of an eccentric disk during superoutbust. (Whitehurst 1988; Hirose & Osaki 1990; Lubow 1991).

Another CV category is the intermediate polars (IP a.k.a. DQ Her systems) (Warner 2003). IPs have similar components as dwarf novae, but the white dwarf has a stronger magnetic field of several MG. Again, material from an orbiting red dwarf forms an accretion disc around the white dwarf. However, the white dwarf's magnetic field is strong enough to truncate the inner portion of the disc. Instead of spinning rapidly through interior portions of the disc and eventually falling onto the white dwarf, accreted material follows the magnetic field lines and falls ballistically onto the two poles of the white dwarf (Hellier 2001). As a result, these systems (such as GK Per) may still undergo disc outbursts but they are rare and of short timescales. IPs also exhibit short-term variability mostly attributed to turbulence in the accretion streams.

BZ UMa was first reported as a variable star in 1968 (Markarian 1968). Classification as a cataclysmic variable was first proposed by Mayall (1972) based on four years of observations in the American Association of Variable Stars Observers (AAVSO) database. Later, its possible nature as a dwarf nova was suggested by Green, et al. (1982) based on spectroscopic observations. Wenzel (1982) suggested that BZ UMa was a hybrid between the UGSU and UGWZ (a.k.a. WZ Sge stars) subclasses of dwarf novae. UGWZ dwarf novae are most notable for very long periods between outbursts, usually on the order of decades. Wenzel based his suggestion on combined data from Sonnenburg plates (Bräuer & Fuhrmann 1992) and the AAVSO that showed few outbursts from 1928 - 1981. However, the average cadence in the combined data sets is one observation per seven days (with an uneven distribution skewed towards the end of the dataset). The typical BZ UMa outburst only lasts four days, so it is possible that many outbursts were not included in the dataset.

Much attention was focused on BZ UMa over the following decades. Skzody & Feinswog (1988) reported infrared variability (possibly due to ellipsoidal variations) and determined an inclination of 60 degrees, which explained a lack of eclipses in the light curve. An orbital period of $P_{orb}= 0.0675d$ was reported by Claudi, Bianchini & Munari

(1990) and refined to $P_{orb}$= 0.0679d by Ringwald & Thorstensen (1990). Jurcevic, Honeycutt, Schlegel & Webbink (1994) determined a mass ratio, $q$=0.20 +/- 0.09 in 192 spectra and 96 *V* band photometric observations over 79 nights. They supported a UGSU classification based on BZ UMa's mass ratio, outburst interval and short orbital period.

Spectral observations have shown a profile typical of a CV with an accretion disk (Szkody, et al. 2003), but with anomalous N/C line ratios (Winter & Sion 2001; Gänsicke, et al., 2003) that could be the result of a secondary star being stripped of its outer layers (Gänsicke, et al. 2003). Anomalous line shapes with as many as five separate peaks in the emission lines have also been reported (Neustroev, Medvedev, Turbin & Borisov, 2002), suggesting the existence of more than one hot spot (Neustroev, Zharikov, Medvedev & Shearer 2004). Neustroev, Charikov & Michel (2006) obtained simultaneous optical photometry and echelle spectroscopy of BZ UMa as it entered an outburst in 2005. They determined that the outburst is of the inside-out variety, which are consistent with low mass transfer rates generally associated with U Gem dwarf novae (Buat-Ménard, Hameury & Lasota 2001; Warner 2003).

Dwarf novae often exhibit short period quasiperiodic oscillations (QPOs) caused by changes in the accretion rate or in the dimensions of intermediate and/or inner areas of the disk (Vrielmann, Ness & Schmitt 2004). They are also common in IPs where they are caused by changes in concentration of material ("blobs") in the accretion stream. QPOs in the BZ UMa system were first reported in Hα radial velocity observations by Ringwald and Thorstensen (1990). They observed long period QPOs of 17.8h and 38.4h over a six day observing run and are the first to suggest the system may be magnetic. Jurcevic, et al. (1994) did not find any QPO activity in their observations. In the 1999 outburst, Kato (1999) detected much shorter QPO activity centered on a period of 0.65h and with an amplitude of 0.3 magnitudes (unfiltered). Subsequently, a QPO with a period of 0.072d ± 0.024 was reported in the 2004 outburst (Price, et al. 2004).

Kato (1999) suggested that these properties, along with the bright spectra lines, X-Ray detection and lack of detected superhumps is evidence for classification as an IP. Ishioka, et al. (2005) analyzed eight outbursts of BZ UMa from 1996 – 2004 and report similarities to outburst behavior very close to that of HT Cam, a confirmed IP (de Martino, et al. 2005). Also, most IPs have bright X-Ray emission and BZ UMa is a bright X-ray source in the ROSAT catalog (1RXS J085343.5+574846; Voges et al. 1999, Voges 2000). Dwarf novae can also have strong X-Ray emission if they are low inclination systems.

Fully magnetic CVs, known as "polars" (a.k.a. AM Her stars), typically have circular polarization <1% (Cropper 1986). IPs are often reported with even less polarization, likely due to a weaker magnetic field and the interference of non-polarized light from the accretion disk. However, detection of polarization is still a useful tool for identifying magnetic influence in a CV (Patterson 1994). An investigation of literature by Butters, et al. (2009) found fifteen IPs with published polarization detections (AE Aqr, AO Psc, BG CMi, DQ Her, EX Hya, FO Aqr, GK Per, PQ Gem, RXJ2133, TV Col, V1223 Sgr, V2306 Cyg, V2400 Oph, V405 Aur & YY Dra). Butters et al. surveyed eight of the objects with new UBVRI photo-polarimetric observations and confirmed the detection of circular polarization in seven of them. In addition to the fifteen quoted by Butters, et al., polarization detections have been reported for the IPs BM CMi (Penning, Schmidt & Liebert 1986) and RXJ1712-2412 (Buckley et al. 1995). Still, the numbers suggest something on the order of 10% of IPs have detectable polarization. BZ UMa was previously observed on March 17, 1980 by Liebert & Stockman (1980). They measured circular polarization during quiescence with a 24-min exposure and a ~3200-8600 angstrom detector on the Kitt Peak 2.3m. They effectively reported a null result below their detection limit of 0.077 +/- 0.17%.

In this paper, we report on analysis of 30 years of visual and photometric observations of BZ UMa along with more resent polarimetry. We combine reports from the literature with our analysis to present a proposed classification.

*Observations*

The AAVSO International Database (AID) (AAVSO 2009) contains a mixture of approximately 17 million visual, photoelectric and CCD observations of ~8,000 variable objects made over the last 115 years. For BZ UMa, it has 12,585 visual observations from 201 observers dating from November 21 1968 to January 1 2009 (Figure 1). Visual observations of variable stars typically have an uncertainty of 0.2-0.3 magnitudes per observation, although some of the more experienced visual observers can be as precise as 0.02 magnitudes (Price, Foster, Skiff & Henden 2007). The AID also has 50,609 CCD observations of the BZ UMa field from 95 observers dating from March 17, 1995 to

December 31, 2008. These observations have a wide variety of uncertainty and typically range from 0.1 to 0.01 magnitudes, although some professional and experienced amateur observers can obtain millimag precision. Observers are provided with a common observing chart to ensure a homogenous set of comparison and check stars. All AAVSO data is available for download from the AAVSO web site. Additional information associated with individual observations is available in the dataset, including photometric properties such as uncertainty, comparison stars used and reports on seeing conditions.

To look for evidence of magnetism, polarimetric observations of BZ UMa were taken on the night of Jan. 17, 2008 (UT) with the Dual-Beam Imaging Polarimeter (DBIP) on the University of Hawaii's 2.2m telescope on Mauna Kea, Hawaii (Masiero, Hodapp, Harrington & Lin 2007). BZ UMa was in a quiescent state at the time. DBIP is an imaging polarimeter capable of simultaneously measuring both linear and circular polarization with precision of better than 0.1% (Masiero, Hodapp, Harrington & Lin 2008). It uses a broadband optical filter with sensitivity from 400 to 700 nm. Each polarization measurement consisted of six images of 5 minutes each to achieve the required polarimetric precision. In total, six sets of measurements were obtained over the course of the night.

*Analysis*

Historical Outbursts

Twenty-nine outbursts are present in the entire AID. Seven outbursts were observed from November 1968 until November 1979. That period has a total of 493 visual observations with an average cadence of around eight days between observations, so it is possible some regular outbursts were missed. However, any superoutbursts should have been detected and reported. No outbursts of any type were observed in a period from November 1979 until April 1991. BZ UMa did brighten above its typical quiescent level of $V$=~16 a few times during this period, however it was never reported brighter than $V$=13.1. Outbursts resumed in April 1991. From then until January 2009 twenty outbursts were detected with an average interval of 308d +/- 105.

Twenty-eight of the twenty-nine outbursts had similar profiles. They lasted less than four days with a rise that takes less than a day followed by a smooth decline over three days. The reported peak magnitude of each outburst ranged from $V$=10.4 to $V$=12.6, with an average peak of $V$=11.2 +/- 0.7 magnitudes. The 2007 outburst was atypical compared to the others (Figure 2). First, the entire outburst lasted roughly 15 days. Second, it had a shallow rise to ~$V$=11 that took about four days, followed by a shallow decline to $V$~=12 over eight days and then a sharp decline to quiescence, $V$~=15, over two more days.

Quasiperiodic Oscillations and Quiescent Activity

To investigate the reported presence of QPOs, the AAVSO organized an intensive 2-day observing campaign on April 17 and April 19, 2004, while BZ UMa was in quiescence (~$V$=16.5). The goal was to minimize gaps in the light curve to control for aliasing in the statistical analysis. Twenty-six observers participated. Enough data was collected to create two continuous light curves of approximately 14 and 16 hours each.

Fourier analysis of the data reveals QPOs with a period around 0.04d on April 17 and QPOs with a period around 0.03d on April 19. The amplitudes of the QPOs are typically a few tenths of a magnitude, but can reach as high as a full magnitude. We also performed Fourier analysis looking for broad profile, short-term periods on photometry of two outbursts in 2005 and one in 2006. We found no evidence of QPO behavior in 2005-2006.

Superhumps

We analyzed CCD observations from the 2007 outburst for evidence of superhumps (Figure 3, Figure 4). To prepare our data for analysis, we combined the *B, V, Rc, Ic* and unfiltered observations into five separate datasets. We included observations from the peak of the outburst (April 14.1, 2007) until it reached quiescence (April 23.5, 2007). Next, we subtracted a linear fit from each of the individual datasets. This served the dual purpose of removing differences caused by the filters and also subtracting the overall trend of the outburst. Then we combined the datasets and used a date-compensated discrete Fourier transform (Ferraz-Mello 1981) with the CLEANEST algorithm (Foster 1995) for period analysis. Results show superhumps with a mean period of $P_{sh}$= 0.07003 +/- 0.00001 days (Figure 5), which is 2.9% longer than the orbital period reported by Ringwald, et al. of $P_{orb}$= 0.0679 days. This is in line with the Patterson (1998) superhump period excess relationship. The first superhump was detected on April 14.1, about six hours before the outburst's peak, and the last superhump was detected on April 23.1, about a day before it began its rapid descent into quiescence. During this time, $P_{sh}$ evolved from 0.0703 +/- 0.0001 to 0.0698 +/- 0.000029 days, which is in line with an expected decrease in period of around 0.6% per magnitude of decline (Patterson, et al. 1993). The mean amplitude of the superhumps was 0.25 magnitudes, slightly below the 0.3-0.4 magnitude typical of superhumps (Warner 2003).

Similar analysis of CCD data from the 2003, 2004, 2005 and 2006 outbursts reveal no superhumps. The mean photometric accuracy of these data sets (1%) is far less than the typical superhump amplitude of UGSU stars (3%). CCD coverage of outbursts prior to 2003 is not consistent enough for a thorough analysis.

Polarimetry

Measured values for the Stokes Q, U, and V vectors as well as the total linear polarization at each measurement and 1-sigma error bars are plotted in Figure 6. We detect neither summed nor time-variant circular polarization at a significant level in the measured wavelengths. Likewise the data do not support any linear polarization above a few-tenths of a percent. Polarimetric observations from DBIP set a strong upper limit for the circular polarization at <0.2%. Bulk linear polarization can be constrained similarly, however time variable linear polarization of <0.3% cannot be ruled out. Optical photometry obtained during the polarimetry observations show no special activity during the observing window.

Discussion

BZ UMa has some characteristics of an IP, but they are not mutually exclusive with a classification as a dwarf nova. First, it is a bright X-ray source, as expected of an IP.

However, dwarf novae also emit X-rays from close to the surface of the star. BZ UMa's inclination of 60 degrees is high, but may not be high enough to completely obscure the surface of the white dwarf. Second, QPOs have been repeatedly detected in both outbursts and during quiescence. Their period of around .03d is in line with those reported for other IPs (ex: GK Per [Morales-Rueda, Still & Roche 1996] and TX Col [Mhlahlo, et al. 2007]), while dwarf novae QPOs tend to be much shorter. However, QPOs are not consistent by their very nature and could be the result of heavy flaring activity. Finally, very weak circular polarization has been detected. A strong detection of circular polarization would have helped the IP case, however no such polarization was detected suggesting the system is not strongly magnetic.

On the other hand, the case for a dwarf novae classification is more consistent. The long term light curve for BZ UMa reflects that of a dwarf novae. When active, its regular outburst cycle mimics that of the SU UMa subclass of the UG class of dwarf novae. However, it once exhibited a long period of quiescence. This is most common with the WZ Sge subclass of UG dwarf novae but is not unheard of with SU UMa subclass stars (Rosenzweig, 2000). Also, its supercycle (time between superoutbursts) is at least >29 years, also similar to WZ Sge stars who have supercycles on the order of decades. The 2007 superoutburst shows clearly defined superhumps. Superhumps are a defining characteristic of both SU UMa and WZ Sge UG stars but are not usually found in IPs (although it has been speculated that QZ Vir could be a superhumping IP [Vrielmann, Ness & Schmitt 2004]).

The zoo of CV classifications is congested and different classes often share properties. For classification purposes, the most important property of CVs is their light curve: in particular the supercycle and presence (or lack thereof) of superhumps. The classification of BZ UMa as a dwarf novae of the U Gem class and SU UMa subclass ia its light curve is also supported by its mass ratio and orbital period.

Summary & Conclusion

In 2007, BZ UMa experienced its first superoutburst on record, complete with superhumps. In addition, new polarization measurements are able to rule out circular polarization above the 0.2% level, suggesting little or no magnetic influence in the system. Combined with analysis of forty years of archival data, we conclude that BZ UMa is not a traditional IP, but instead it is a member of the U Gem class, SU UMa subclass of dwarf novae. The system still has many unexplained peculiarities (supercycle variability, QPOs, significant flaring in quiescence, and abnormal emission lines). X-ray determination of the white dwarf spin period would be very helpful to further define this enigmatic star system.

Acknowledgements

We would like to thank Paula Szkody for advice given subsequent to a talk on this topic. We would also like to thank the following AAVSO observers for additional contributions to the paper: R. Huziak, R. McDaniel, J. Ott, N. Quinn, G. Walker. This research has made use of the SIMBAD database, operated at CDS, Strasbourg, France and the International Variable Star Index (VSX), operated by the AAVSO.

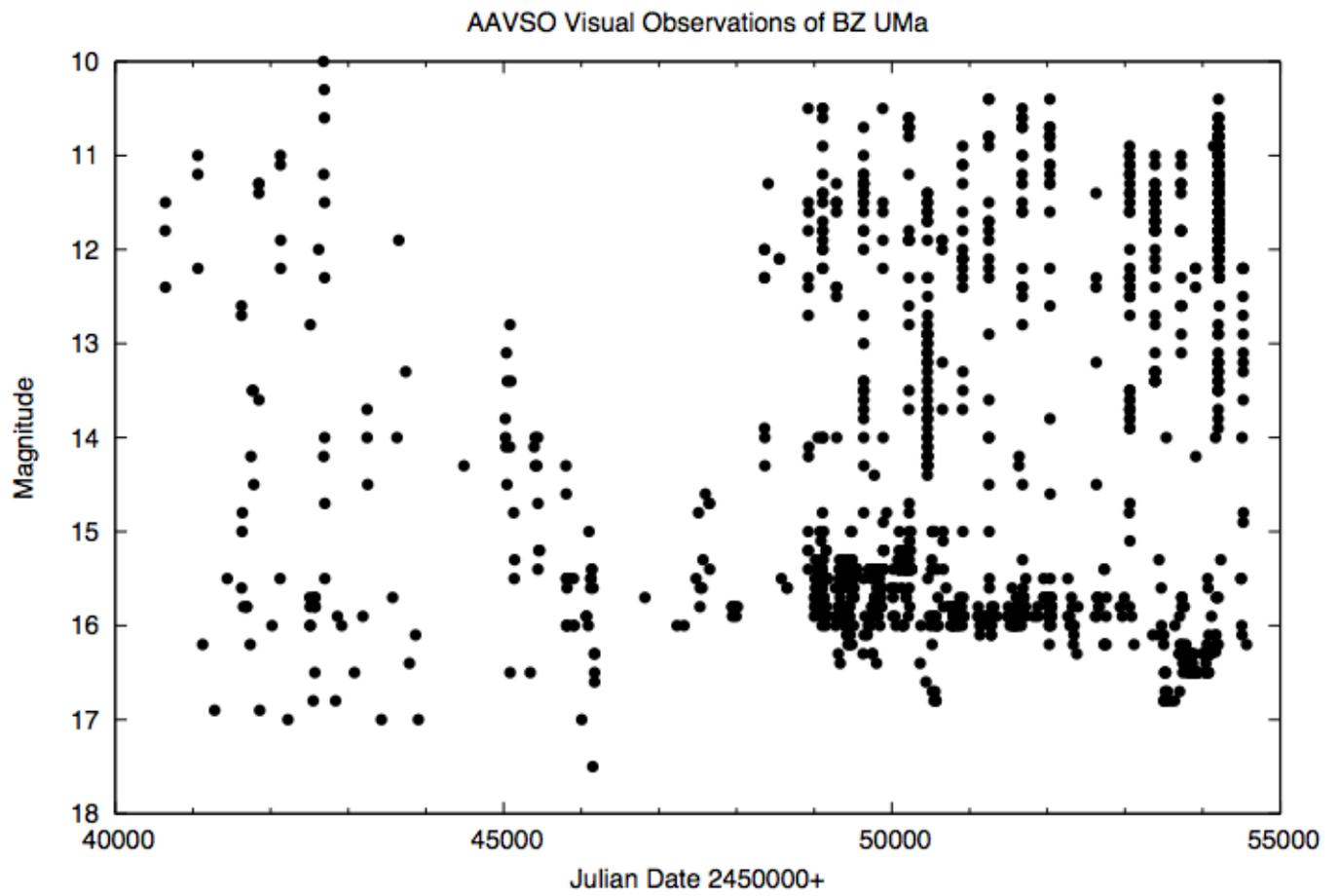

Figure 1. Visual positive observations of BZ UMa from November, 1968 – January, 2009.

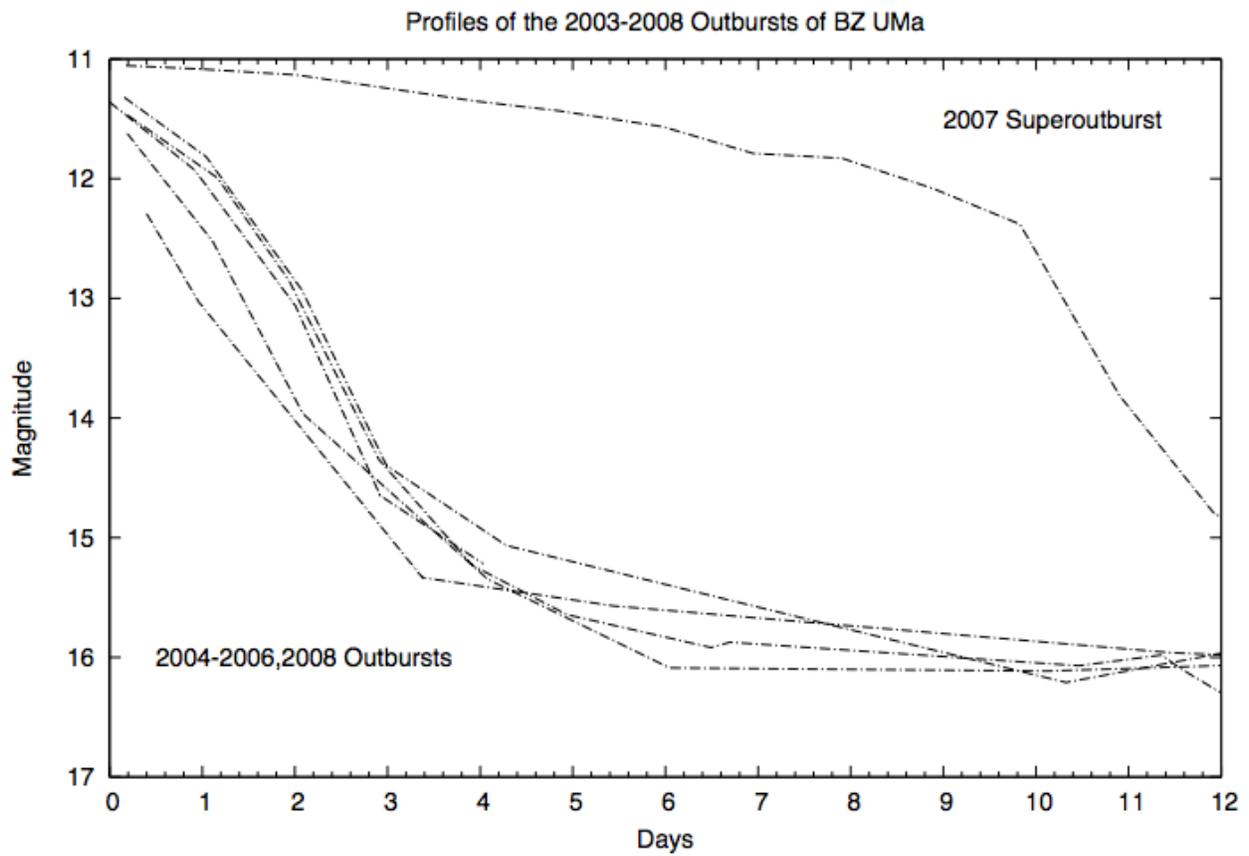

Figure 2. One day mean curves of the 2003-2008 outbursts of BZ UMa synchronized on the peak time of each outburst.

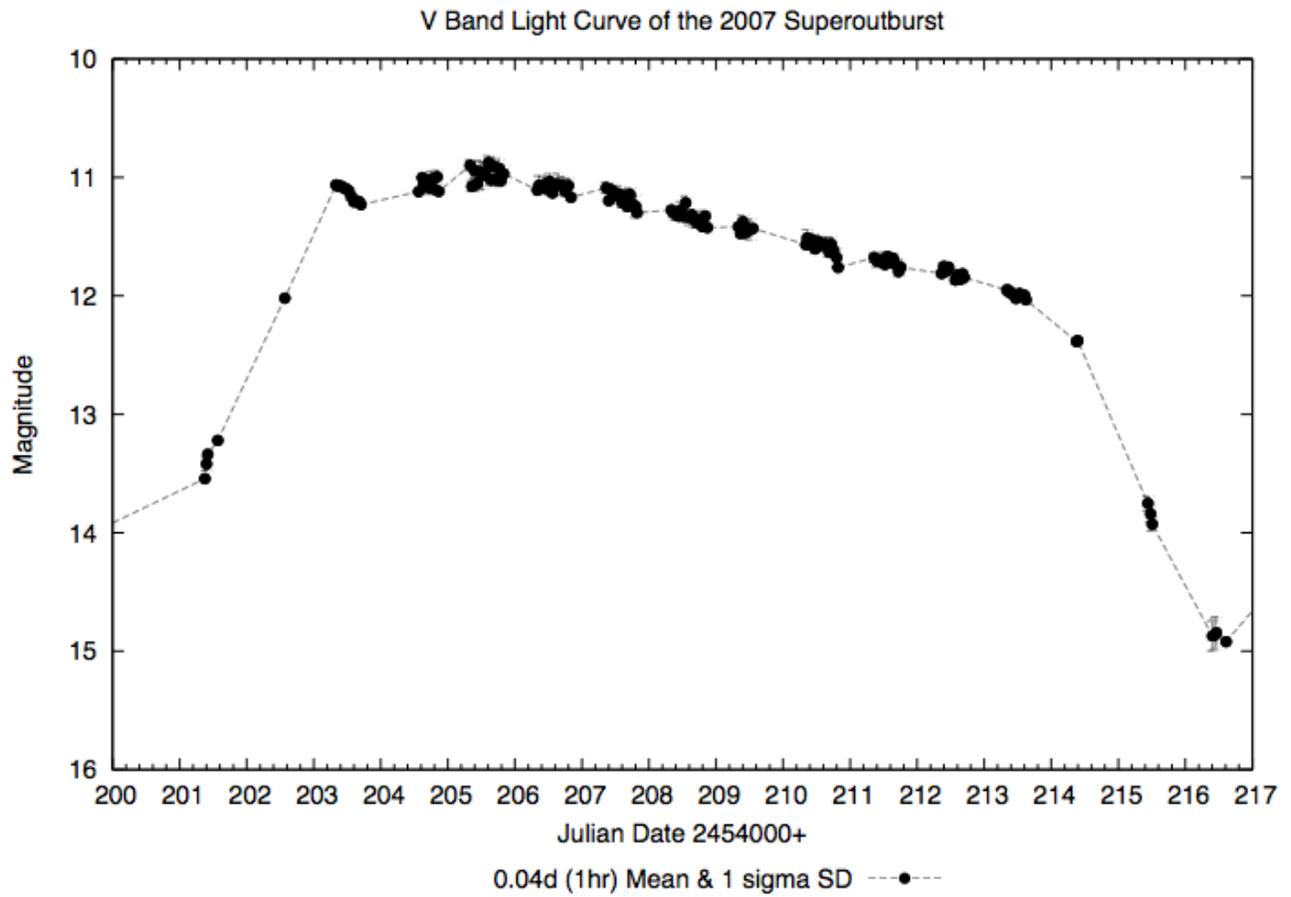

Figure 3. CCD *V* band observations of the 2007 superoutburst binned into 0.04d intervals with 1 sigma standard deviation.

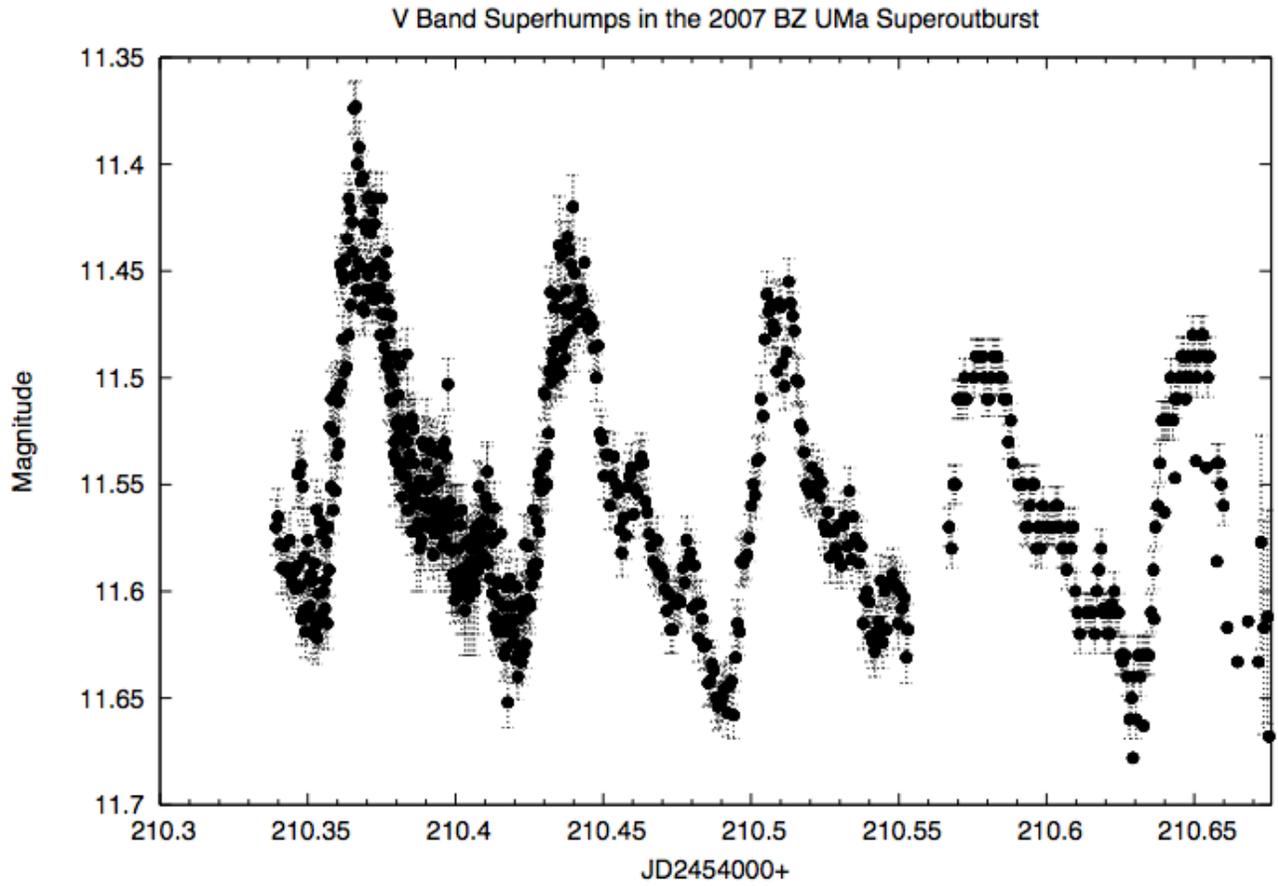

Figure 4. Sample *V*-band light curve of superhumps in the 2007 superoutburst. Observations in this light curve were contributed by David Boyd, Keith Graham, Jeremy Shears, Ray Tomlin and Arranz Teófilo.

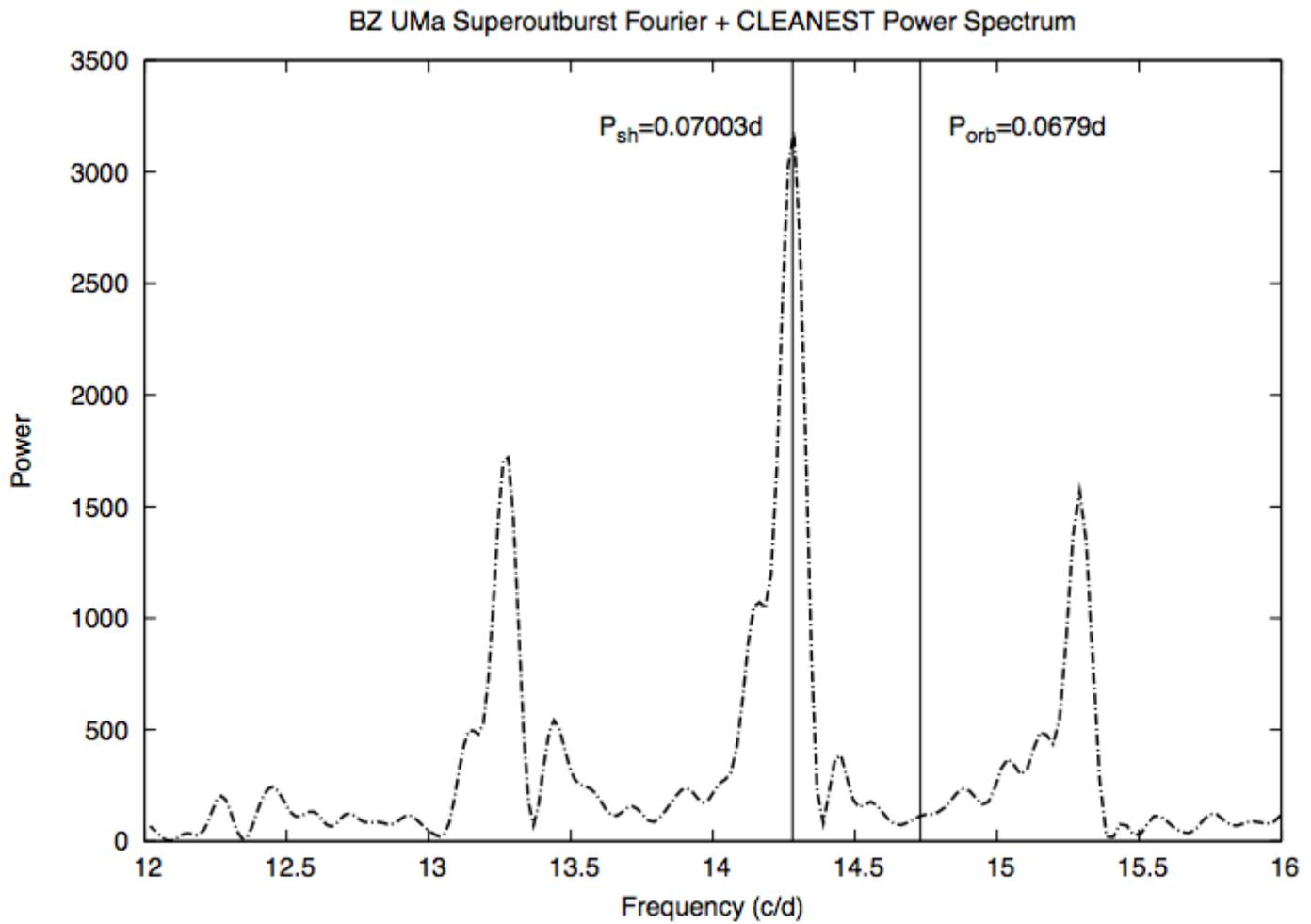

Figure 5. Fourier plus CLEANEST analysis of detrended CCD *B*, *V*, *Rc*, *Ic* and clear band observations from the 2007 outburst. The peaks on either side of the main 14.3 c/d peak are sampling aliases at +/- 1 cycles per day.

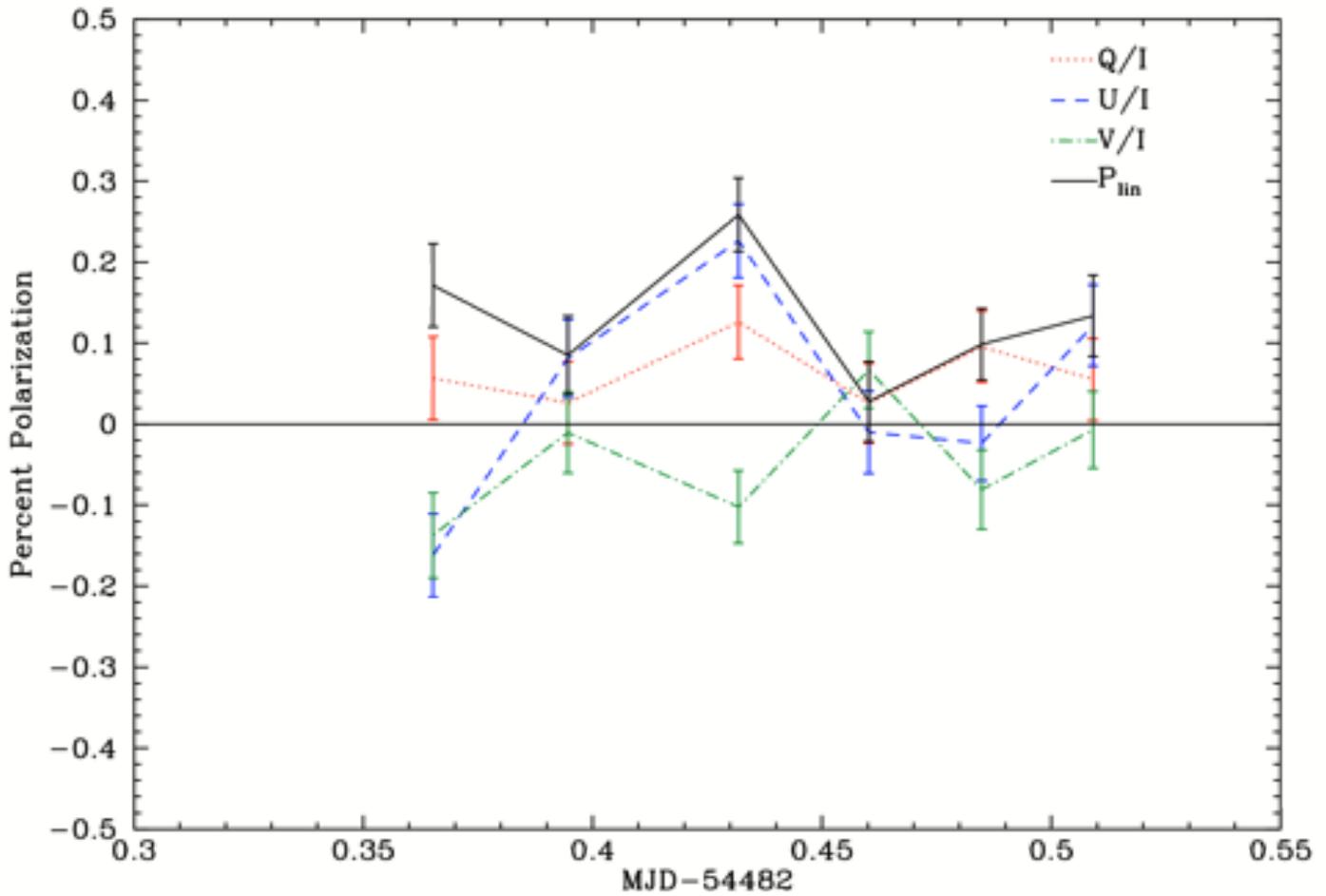

Figure 6. Polarimetric observations of BZ UMa made with the Dual-Beam Imaging Polarimeter (DBIP) on the University of Hawaii's 2.2m telescope on Mauna Kea, Hawaii. Measured values for the Stokes Q, U, and V vectors as well as the total linear polarization at each measurement and 1-sigma error bars are plotted.